# Wrong-way risk in credit and funding valuation adjustments

*Mihail Turlakov*[1]

**Abstract**

Wrong-way risk in counterparty and funding exposures is most dramatic in the situations of systemic crises and tails events. A consistent model of wrong-way risk (WWR) is developed here with the probability-weighted addition of tail events to the calculation of credit valuation and funding valuation adjustments (CVA and FVA). This new practical model quantifies the tail risks in the pricing of CVA and FVA of derivatives and does not rely on a limited concept of linear correlation frequently used in many models. The application of the model is illustrated with practical examples of WWR arising in the case of a sovereign default for the most common interest-rate and foreign exchange derivatives.

-

Counterparty and funding risks in financial markets has been attracting much attention especially since the beginning of 2007-2008 global financial crisis. Both types of risks became major factors in the upheaval of quantitative finance with far reaching implications for the valuation and risk-management of financial derivatives. Credit Valuation Adjustment (CVA) and Funding Valuation Adjustment (FVA) to the price of a derivative are the manifestations of this upheaval as well as various corresponding regulatory measures such as CVA VAR (capital charge for CVA volatility) and liquidity ratios imposed on banks by the Basel 3 accord.

The most dangerous manifestation of counterparty risk is the so-called wrong-way risk (WWR) – positive correlation[2] between the counterparty's default and the exposure (mark-to-market or replacement value) - can be particularly harmful to the stability of the financial system and each individual bank's balance sheet during extreme market dislocations (also known as tail events). Due to the interconnectedness of the financial system, major systemic or economic shocks triggered are greatly amplified by potential counterparty related losses and funding stresses, and especially so if further magnified by WWR. Importantly, the total large exposures of WWR type are most likely the consequence of a financial institution's business model due to concentration, leverage and funding risks and therefore deserve a close attention and active management on all levels.

In the context of counterparty risk or CVA risk-management, the "positive tendency" between the exposure and the likelihood of default is frequently expressed as a correlation between corresponding market factors and the counterparty's default intensity. Several models were developed in the literature based on Merton's bankruptcy model (Levy 1999) (Buckley 2011), stochastic dynamic credit models (Finger 2000) (Redon 2006), parametric dependency (Hull 2012) and Gaussian copula (G. Cespedes J.C. 2010). While these models are helpful in pricing and "daily" risk-managing of cross-gamma due to credit and market factor (as FX, rates, etc) correlations, these models are not reliable and even misleading in the cases of tail events and large economic cycle swings of multiple market asset classes involved in the CVA calculation. Practitioners tend to think in terms of scenarios and to be particularly concerned about WWR of tail events (Pengelley 2010), and from this point of view we develop a simple model which expresses the tail-risk events explicitly and therefore intuitively. *The main challenge of WWR is to identify and to characterise various extreme WWR scenarios* in order to be able to risk-manage and perhaps to steer or diversify the business away from WWR dangers; the presented tail-risk model helps to analyse WWR and is designed to be practical and "as simple as possible yet no simpler".

---
[1] CVA Portfolio Management, WestLB, London. The views expressed are the opinions of the individual author only. The author received no financial support for this paper and he declares no competing financial interests. No opinion given in this article constitutes a recommendation by the author.
[2] More generally, the positive tendency of both quantities - the exposure and the probability of default - to increase during certain period but not necessarily in the sense of stable correlated, daily and no-lag, changes.

**The motivation and the background**. The full risk-management solution for the calculation of counterparty's exposures and CVA reserves presents a formidable engineering and modelling challenge for financial institutions. A cross-asset portfolio dependent on hundreds of market factors requires a modular and scalable platform, and WWR is one of the last and complex requirements in the design of such a platform. In the backdrop of this complexity, the advantage of the proposed tail-risk model is that it can be conveniently developed as a modular "after-model" based on the base model by consistently incorporating tail-risk scenarios.

The reliance on cross-asset correlations in the context of large CVA portfolios can be questioned. Not only these correlations can hardly be calibrated but, more importantly, this type of correlations is very unstable and variable[3]. Correlations can help to skew the centre of the distribution but they cannot model well the tails[4]. In addition to the various technical reasons, a fundamental reason to go beyond the conventional no-lag linear correlations is due to the nature of long-period (over and beyond 1-3 years) economic business cycles and trends as well as idiosyncratic[5] market reactions if seen across multiple asset classes. In theory and frequently in the markets reality, the growth and recession cycles are abated by central banks with rate hikes and cuts which are followed in turn by the corresponding changes in the demand for credit in the economy. These considerations lead to positive co-tendency between credit spreads and government rates observed on long-time scale of 1-2 years, while no-lag correlation, in fact, changes the sign depending on the stage of the economic cycle (Lorenzen 2011), see Fig.1. In the context of WWR modelling and risk-management, the institution must be as much concerned with large swings of economic cycles accompanied by possible actual defaults in concentrated exposures as with the daily volatility of CVA reserves due to daily correlations.

---

[3] The analysis of various credit-FX, credit-rates, etc shows simply no definitive linear no-lag correlations over extended periods of time. The uncertainty of cross-asset correlations can be contrasted with other types of correlations which are more reliable and reinforced by the market for various reasons. For instance, FX-FX correlations between two FX rates with mutual currency are very natural and stable. Another example of a more persistent correlation which describes the skew volatility effect is the correlation between the underlying security and its volatility.

[4] Correlation-based models may not have fat tails in their distributions at all, or, at most, a fat tail on the one side of the distribution, while tail-risk approach can easily express fat tails on both sides of the distribution which is the reality of some markets. Furthermore, the effect of correlations in the tails of the distribution necessarily simulated via Monte Carlo in portfolio CVA calculations is typically very noisy due to insufficient averaging.

[5] For instance, dependent on the nature and timing of the governments' interventions.

**The model**. We consider a single stressed scenario of WWR added to the base model to explain the essence of the model, and we will generalise to multiple stress scenarios afterwards.

One major stress scenario is the sovereign's default accompanied by various market dislocations. In general, *the main assumption of the model is that various stress scenarios (sovereign default, commodities shock, etc) can be defined in terms of stressed macroeconomic parameters (stressed rates and discount curves, FX rates, commodity and housing prices, etc) and the probabilities of those stress scenarios.* In our example of a sovereign default, the probability of the default can be implied from the sovereign CDS level.

The expected positive exposure (EPE) which enters into the CVA calculation (Gregory 2012) is always meant to be the conditional exposure given the default of the counterparty. Since the major dislocation and associated WWR occur only when the sovereign defaults, the total EPE is the probabilistic sum

$$EPE_{WWR} = P(sov|Cpty) * EPE^{stressed} + \left(1 - P(sov|Cpty)\right) * EPE \quad (1)$$

where $P(sov|Cpty)$ is the conditional probability of sovereign default given counterparty's default, $EPE^{stressed}$ is EPE calculated under a stressed market, while $EPE$ is calculated from the base model under the current market conditions. If the counterparty is the sovereign itself, then $EPE_{WWR} = EPE^{stressed}$ and the reasoning is exactly the same as in the reference (Levy 1999) in the case of a sovereign default. Note that in a crisis some of the market dislocations can be already expressed (priced in), so EPE will contain some or all effects of the stressed $EPE^{stressed}$. To avoid this difficulty, we suggest a simple interpolation which smoothly connects the unstressed and stressed limits[6]:

$$EPE^{stressed} = EPE * \tanh\left(\frac{P(sov)}{P_{threshold}}\right) + EPE^{abs-stress}\left(1 - \tanh\left(\frac{P(sov)}{P_{threshold}}\right)\right), \quad (2)$$

where $EPE$ is still the current market EPE, $EPE^{abs-stress}$ is EPE calculated with the market parameters stressed to certain absolute levels. While $P_{threshold}$ is the threshold default probability which is defined as near-default or half-life threshold and marks the crossover between crisis (stressed scenario) and non-crisis limits. For instance, in our example, this threshold could correspond to sovereign CDS level of 1000bps.

We can define an effective parameter $\lambda$ relating unconditional and conditional probabilities $P(sov|cpty_i) = \lambda_{cpty_i} P(sov)$ in order to introduce relative strength of "the coupling"

---

[6] The aim of the interpolating function $\tanh(x)$ is to have a sharp transition between two limiting cases $\tanh(x \ll 1) \approx 0$ (unstressed limit $P(sov)/P\_threshold \ll 1$) and $\tanh(x \gg 1) \approx 1$ (stressed limit $P(sov)/P\_threshold \gg 1$).

between different counterparties and their sovereign. The meaning of the coupling parameter $\lambda$ is transparent; the larger $\lambda$ corresponds to larger systematic relevance of the counterparty in the sense of coupling of the fates of the counterparty and its sovereign. For a major bank, this parameter may be large (certainly larger than 1, but still $\lambda P(sov) \leq 1$) or small very much dependent on the specific country's resolution of a default. If a company's default is quite independent from the sovereign, this parameter is 1. In a special and unusual situation, if sovereign's and counterparty's default are mutually exclusive, then the parameter is zero.

Combining all the above equations, we derive a final expression which expresses clearly the correction due to a stressed WWR scenario

$$EPE_{WWR} = EPE + \lambda P(sov)(EPE^{abs-stress} - EPE)\left(1 - \tanh\left(\frac{P(sov)}{P_{threshold}}\right)\right). \qquad (3)$$

The simplicity of the model is underlined by a transparent meaning of all the parameters involved. Interestingly, WWR in the model is dynamic because $P(sov)$ is the current-market implied probability from the market-based CDS. In this sense the model also contains the correlation effect between sovereign's and counterparty's CDSs since the full CVA expression is the product of counterparty's default probability (and therefore its CDS) and $EPE_{WWR}$ which contains sovereign's $P(sov)$ in turn.

**Selected results of the model's application.** In this section, we illustrate the application of the model by considering some liquid contracts. We start with an interest rate swap (IRS) of 10 year maturity; the counterparty has CDS=300bps (flat curve, for simplicity), and the sovereign has CDS=200bps (for instance, one of the core countries in Europe). The interest rates are low with the upward slope across the maturities (as in April 2012). We specify a tail-event scenario for IRS by assuming that the rates will move by 500bps across the curve in the case of sovereign default. The only other assumption which we need to make is that the coupling parameter $\lambda$ is equal to 1 corresponding to "independent company" coupling strength.

Without any WWR, CVA for payer IRS (paying fixed rate and receiving floating one) is equal to 7.8bps (or 0.078%) running. The payer IRS has WWR because our exposure on the swap increases if floating rate increases. With WWR CVA for payer IRS becomes 12.7bps which is significantly higher than 7.8bps. Fig.2 illustrates various EPE profiles for payer IRS, which are the stressed exposure $EPE^{abs-stress}$, the exposure $EPE$ without WWR, and the total

exposure $EPE_{WWR}$ as well as forward mark-to-market (MtM). For comparison, the receiver

IRS has a right-way risk (RWR) in this stressed scenario, and CVA without and with RWR is 2.8bps and 2.4bps correspondingly. On the portfolio level, the imbalance between WWR and RWR is mainly determined by the nature of bank's portfolio (usually, receivers make up most of derivatives' exposure with uncollateralised corporate counterparties) and the types and probabilities of stressed scenarios.

Another clear-cut example of WWR is the cross-currency swap (CCS) when in the final exchange of notionals we receive a notional in an external currency (i.e. USD) and pay in the country's own currency (i.e. BRL) and the counterparty's business is in the same emerging country (i.e. Brazil). In the scenario of a sovereign default, the emerging (BRL) currency is devalued and therefore our exposure spikes, while the credit spread of the counterparty will almost certainly widen, this is WWR. The cases of sovereign defaults in emerging countries and accompanied currency devaluations are very well-documents (South-East Asia crisis of 1998, Russia, Argentina, etc).

Let us consider the following assumptions: CCS of 5 year maturity, counterparty's CDS=500bps, sovereign CDS=130bps, again $\lambda = 1$, and the amount of the devaluation[7] of BRL currency in the default is estimated to be 40%. The calculation based on the tail-risk model shows that CVA increases from 82bps to 87bps on the account of WWR.

Since the stressed tail events are considered explicitly, the tail-risk model for CVA can be quite easily applied to any contracts or exposures including such important cases as CDS and a portfolio's exposure under the collateralisation. In the case of collateralised exposures, bi-lateral or via forthcoming central counterparties (CCPs), WWR in FVA is even more important, and this is what we consider next.

**WWR in funding[8].** WWR in funding of derivatives positions is very critical due to frequent (daily, or, even, throughout the day in the case of CCPs) liquidity demands in the posting of the collateral. In the context of the funding, WWR is the "positive tendency" between an institution's funding spread $FS_{t_i}$ and the collateral it is required to post against mark-to-market (MtM) positions. For instance, in the case of a sovereign's default shock, major banks of the country will face significantly higher funding spreads and may also need to post large amounts of collateral to other counterparties due to sudden change of interest rates and FX rates (for example, on a portfolio of CCSs). Without going into the complexity of collateralised trades pricing, we can derive the following expressions for the symmetric FVA in the presence of WWR in somewhat schematic form:

$$FVA_{WWR} = FVA - \lambda \sum_{t_i=0}^{T} (t_{i+1} - t_i) P_{t_i}(sov) \left( FS_{t_i}^{stressed} * MtM_{t_i}^{stressed} - FS_{t_i} * MtM_{t_i} \right) IF_{t_i}, \quad (4)$$

where $FS_{t_i}$ is the funding spread at some forward time $t_i$, $MtM_{t_i}$ is the expected discounted forward MtM, $FS_{t_i}^{stressed}$ and $MtM_{t_i}^{stressed}$ are the funding spread and forward MtM under stressed scenario, and $T$ is the longest exposure date of the portfolio. The collateral required for posting is $-MtM^{stressed}$, and the funding spread cost is paid on this amount. The current FVA is $FVA = \sum_{t_i=0}^{T}(t_{i+1} - t_i)(-FS_{t_i} * MtM_{t_i})$. Similarly to the considerations above, we need to have an interpolation factor in Equation (3), for instance in the form used above $IF = 1 - \tanh(P(sov)/P_{threshold})$.

---

[7] The jump amount of FX devaluation can be implied from the so-called quanto CDS, a spread between CDS traded in home (internal) and hard (external) currencies, see for the detailed discussion (Ehlers 2006).
[8] For a consistent treatment of CVA, DVA, and FVA, see for instance the papers (Burgard 2011) (Morini 2011) and references therein. The distinction between unique bi-laterally agreeable price and economic value for either side of the transaction is even more pertinent in the context of WWR.

To give a practical example, in the case of 10 year IRS payer considered above FVA= -2.55bps with flat funding spread of 100bps. In the case of the above scenario of sovereign's default and the institution's funding spread widening to 200bps FVA becomes -2.96bps.

**Calibration.** The meaningful calibration of any model for a large multiple-asset CVA book which includes WWR is challenging if not nearly impossible task. The tail-risk model needs the specification of the state of the world for each stressed WWR scenario. In the first instance of the model's application, a trader or a risk-manager can specify the parameters of a WWR scenario based on experience and selective historical data, as demonstrated in the examples above. In particular, the counterparties can be classified into several broad WWR ratings, in analogy to credit ratings, and ascribed $\lambda = 1, 10, 100$ (for low, medium, and high WWR) [9]. After calibration, as a sanity check, the results of the tail-risk model can be assessed in comparison with correlation-based models. For instance, in the above example of 5-year USDBRL CCS, a correlation model has qualitatively similar increase in the EPE profile[10] due to WWR with the correlation of order of 20-40% between a hazard rate and FX rate.

The generalisation for multiple independent and "orthogonal" stress scenarios is in principle straightforward where the same correction is summed over all tail events in Equation (3). The total probability of N tail events $\sum_{i=1}^{N} \lambda_i P_i(t)$ should be kept low since the tail events are rare.

While in the context of the sovereign default we can infer the market-implied probability of this tail event from sovereign CDS, for other types of shocks it is necessary to be imaginative yet practical to infer appropriate probabilities[11]. In general, the tail events can overlap and are not independent, and a coherent Bayesian net of the events needs to be constructed. The coherent stress testing and the importance of the tail-risk modelling are discussed in a recent book (Rebonato 2010), and WWR is a very natural context for the application of the tail-risk modelling and Bayesian approach. The importance of the focused modelling of WWR in the situation of the systemic contagion - a type of tail event - by models with FX jumps and increased volatility has been raised recently (Sokol 2012).

---

[9] By applying Bayes' theorem, the coupling λ=P(Cpty│sov)/P(Cpty) =P(sov│Cpty)/P(sov) can be also estimated from historical data as either of the ratios of the probabilities. From the hedging perspective, we are interested in market-implied default probabilities rather than actual default probabilities (for the review, see (Giesecke 2011). Market-based conditional and unconditional proxy-default probabilities can be estimated as the proportion of days counterparty's CDS exceeded the threshold during a certain time period by applying the threshold condition used in Equation (2).

[10] In general, the tail-risk model and correlation models are of very different nature and their results do not need to be similar in any way.

[11] For instance, in the case of commodity shock the implied probabilities of large market moves can be implied from corresponding derivatives markets (i.e. one-touch or digital options).

**The discussion and the summary.** The hedging of WWR is particularly important in the case of concentrated exposures across many counterparties; WWR exposure in the sovereign's default example can be hedged with sovereign CDS and is proportional to $\sum_{i=1}^{counterparties} \lambda_i P_i(cpty)\left(EPE_i^{rel-stress} - EPE_i\right)$. Practitioners already use essentially the tail-risk style analysis of the counterparty risk at several different banks to the author's knowledge due to the simplicity and the practicality of the model. Stress testing is also now widely performed by regulators and by banks as a measure of the stability of the system as a whole and for each individual bank, and the stress testing and WWR modelling can be naturally integrated in the framework of the model.

In conclusion, the presented tail-risk model can be used for the quantification and hedging of WWR in the context of counterparty and funding exposures as well a for capital and stress-testing requirements based on the same $EPE_{WWR}$ calculation. The methodical recognition of tail-events WWR and corresponding practical Bayesian calibration of such a model will help financial institutions not only to protect themselves from extreme events and deep cyclical economic stresses but also to benefit them by understanding the important WWR part of the economic value of counterparty and funding risks[12].

---

[12] The author would like to thank D. Pfeffer, J. Keenan, and A. Farmafarmaian for their thoughtful comments which greatly improved the article.

Graph 1. The history of US Treasury 10Y yields and corporate spreads based on Moody's Bond Corporate BAA Index (in % percentage value). Different periods can be seen with positive and negative co-tendency between government yields and corporate spreads.

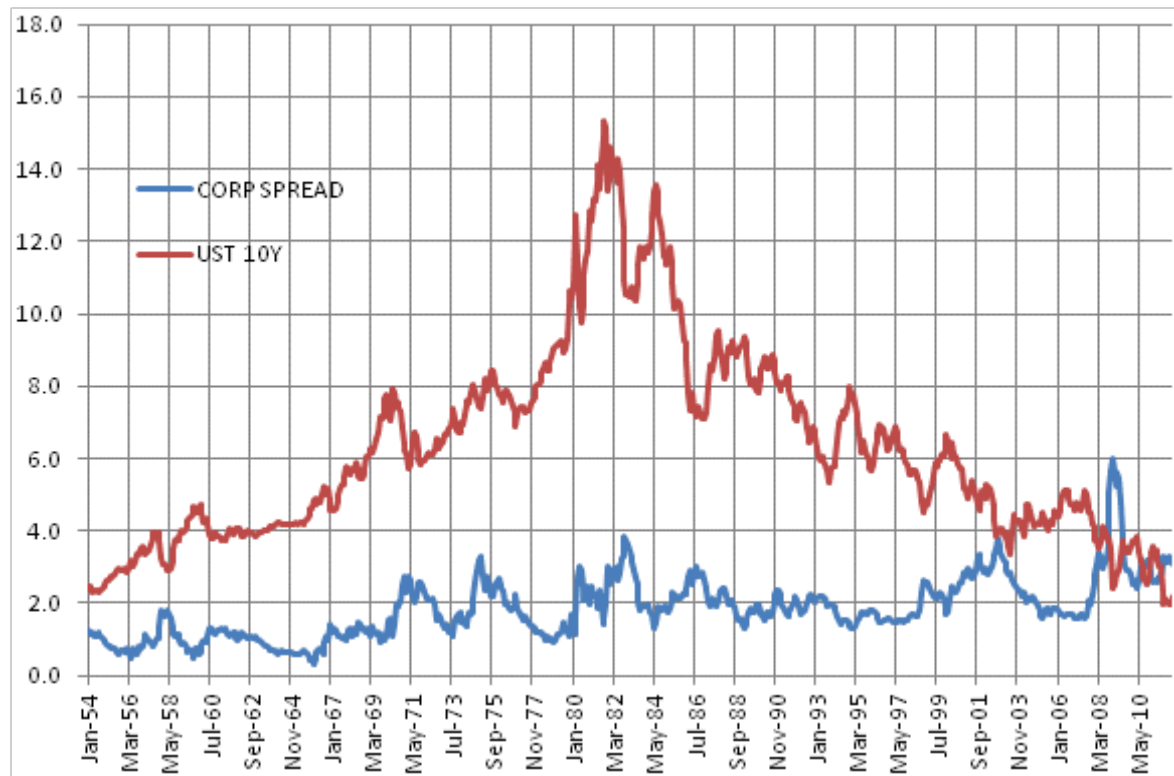

Graph2. Various exposures for a payer IRS. The stressed exposure is the case of rates jumping by 500bps. The magnitude of EPE is given as a percentage of IRS notional.

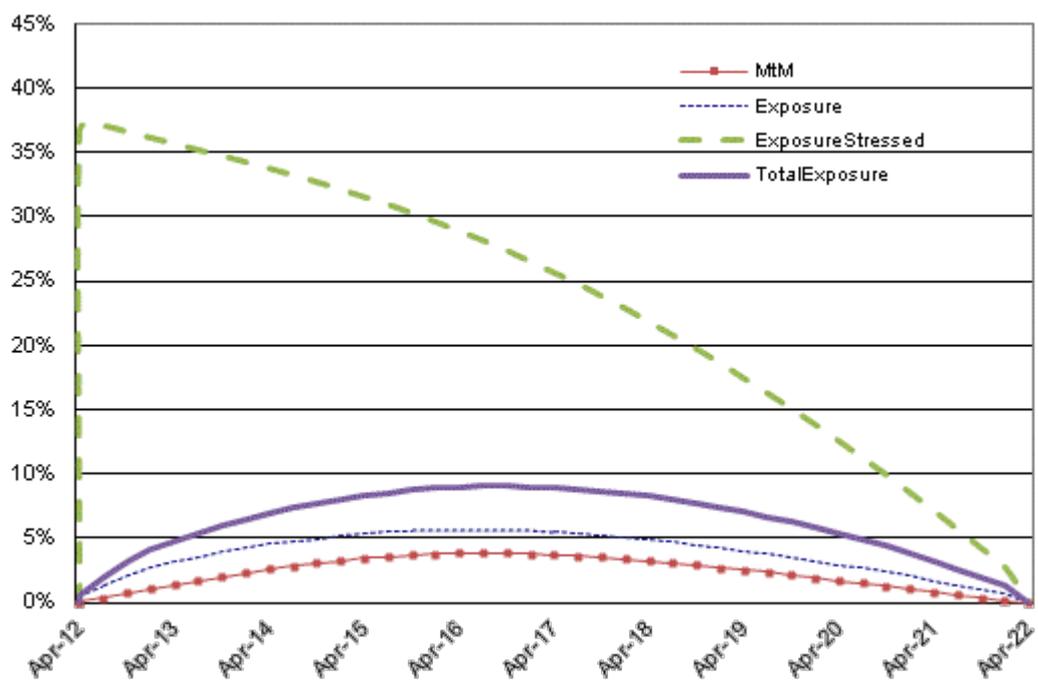